\documentclass[twocolumn,showpacs,aps,pra,amsmath,amssymb]{revtex4}
\usepackage{graphicx}
\usepackage{dcolumn}
\usepackage{bm}
\usepackage{times}
\usepackage{epstopdf}

\usepackage{amsmath,epsfig}

\newcommand{\ket}[1]{\left\vert#1\right\rangle}

\begin{document}
\author{Mauro Paternostro\mbox{$^{1}$}}
\author{Hyunseok Jeong\mbox{$^{2}$}}
\affiliation{ \mbox{$^{1}$}  School of Mathematics and Physics, Queen's
University, Belfast BT7 1NN, United Kingdom\\
$^2$ Center for Subwavelength Optics, Department of Physics and Astronomy, Seoul National University, Seoul, 151-742, Republic of Korea}
\date{\today}
\begin{abstract}
We investigate the violation of non-local realism using entangled coherent states (ECS) under nonlinear operations and homodyne measurements. We address recently proposed Leggett-type inequalities, including a class of optimized incompatibility inequalities proposed by Branciard {\it et al.}, Nature Phys. {\bf 4}, 681 (2008) and thoroughly assess the effects of detection inefficiency.
\end{abstract}


\title{Testing non-local realism with entangled coherent states}

\maketitle

Correlations among systems are important in modern physical science. They frequently allow us to unveil hidden aspects of natural phenomena and,
remarkably, they represent a powerful {\it litmus test} 
in the study of the differences between classical and quantum worlds.
In fact, quantum mechanics allows correlations which have no counterpart in the classical domain and thus represent the 
intrinsic advantage opon which some applications of quantum information 
processing are based~\cite{horo}. 

The concepts of {\it entanglement} and {\it non-locality}~\cite{bell} embody the most striking examples of the profound implications of quantum correlations in the behavior of 
multipartite systems~\cite{bell}. The work by Bell in this respect is a milestone in providing a 
fundamental {test} for proving how the (intuitively reasonable) joint 
assumptions of locality and realism are in striking contrast with the 
description of quantum mechanical correlations~\cite{bell,exp}. The enormous 
interest given to investigations around Bell's inequality in the last thirty 
years, however, has not yet clarified in an unambiguous way the interplay between the two assumptions. In this context, the proposal by Leggett for a non-local 
realistic model stands as a seminal contribution~\cite{leggett}, which has encountered a quickly-growing interest by the physics community at both the theoretical and experimental level. The original idea by Leggett has been recently put within the grasp of current state-of-the-art experimental capabilities by a clever re-formulation of his incompatibility theorem~\cite{vienna1}. Some of the most demanding assumptions behind the formalism in the latter work have been subsequently relaxed in a way so as to make the experimental test of non-local realism more experimentally-friendly. In particular, the requirement for rotational-invariance of the correlation function entering Leggett's inequality can be successfully bypassed~\cite{vienna2,singapore1,singapore2}. The efforts conducted so far have almost exclusively involved linear-optics settings where bi-photon entangled states generated via parametric down conversion have been used as resources for testing non-local realistic assumptions~\cite{vienna1,vienna2,singapore1,singapore2,migdall}. In these cases, the probed non-classical correlations were encoded in the discrete-variables embodied by photonic polarization degrees of freedom.  

However, it has long been known that quantum correlations encoded into states of continuous variables (CV) can violate Bell's inequalities: among others, Banaszek and W\'odkiewicz have proven that Bell's inequality as formulated by Clauser-Horne-Shimony-Holt (Bell-CHSH) can be violated by Gaussian CV states upon parity measurements and displacement operations~\cite{bw} and Chen {\it et al.} have identified a pseudo-spin formalism which optimizes the Bell-CHSH inequality violation~\cite{chen}. Jeong {\it et al.} have studied Bell's inequality tests for CV states with dichotomic observables~\cite{jacobmyungetal} while Paternostro {\it et al.} have addressed the case of Gaussian CV states measured by standard homodyne detectors~\cite{ioejacob2}.

Remarkable examples of CV resources are provided by entangled states of two quasi-distinguishable coherent states, or entangled coherent states (ECSs)~\cite{ECS}, which are useful resources in many quantum information processing tasks. While, for the sake of conciseness, we omit a discussion on the ample range of applications that ECSs have found in these years~\cite{examples,jacobmyung}, it is important to mention that Bell's inequality violation with ECSs and their mixtures has been successfully investigated using, for instance, photon-parity measurements and dichotomic measurements~\cite{derek,jacobmyungetal,jacobralph}. More recently, it has been shown that even threshold detectors or {\it classical} measurements such as homodyning (which are both unable to reveal single quanta) can be used for a Bell-CHSH test, when an appropriate set of local operations is available~\cite{stobinska}. This approach has been useful in the demonstration of non-locality properties of highly mixed states close to the classical border, as in Ref.~\cite{ioejacob}, where it was shown that even extremely inefficient measurements in the classical limit may be used to demonstrate significant violation of local realism. 

Guided by the success in revealing Bell's inequality violations with ECSs, here we investigate non-local realism of an ECS through local nonlinear operations and homodyne detection and prove that the violation of Leggett-type inequalities is not an exclusive privilege of discrete-variable quantum correlated systems~\cite{ioejacob}. We develop a formal apparatus for the determination of the proper joint correlations entering Leggett-type functions proposed in recent formulations of inequalities showing the incompatibility of non-local realism and quantum mechanics. Remarkably, we demonstrate that for an ECS having large enough amplitudes of its coherent state components, which makes them explicitly multi-photon, the degree of violation of such inequalities becomes optimal and the associated Leggett functions mimic the behavior expected for two-qubit singlet states. We also study the effects of homodyne detection inefficiencies and highlight a strategy to counteract them. Differently from any test performed with bi-photon states, our proposal allows to compensate the spoiling effects of detection inefficiencies simply by preparing an appropriate ECS resource, which can be done off-line. Moreover, an interesting comparison between Bell and Leggett functions against the amplitude of an ECS is revealed, which is a unique feature of our study. Although the experiment proposed here presents some challenges, its experimental realization is not far fetched. In fact, we believe our study will provide additional motivations towards the achievement of large nonlinear effects in quantum optical devices for tests of fundamental physics and the processing of quantum information.

The remainder of this paper is organized as follows. In Sec.~\ref{tools} {we first briefly discuss the basic assumptions in Leggett's original model~\cite{leggett} and concisely discuss their implications for non-local correlations. We then introduce the entangled resource we use throughout our study, the local unitary operations that Alice and Bob should implement and obtain an explicit form for the universal correlation function of the outcomes associated with bilocal homodyne measurements.} This will be the building block for the analysis performed in Sec.~\ref{viennaformulation}, where the simplest of the Leggett-type inequalities proposed in Refs.~\cite{vienna2,singapore1} is studied. Sec.~\ref{singaporeformulation} addresses the case of a recently derived optimal Leggett-type inequality, which is quantitatively studied and compared to the case of Sec.~\ref{viennaformulation}. In Sec.~\ref{inefficiencies} we account for the effects of detection inefficiency, showing that a strategy exists for effectively counteracting such non-ideal experimental conditions. Finally, Sec.~\ref{conclusions} summarizes our findings and presents a brief discussion on issues of practical feasibility of the proposed experiment. 

\section{Resource, tools and general formalism}
\label{tools}

\subsection{Brief summary of Leggett's inequality}

{The model introduced by Leggett in his 2003 paper~\cite{leggett} follows other investigations aiming at identifying the fundamental features that define quantum mechanics. 
Given the state of a bipartite system (in~\cite{leggett} this was encoded in the polarization degrees of freedom of bi-photon states), the crucial assumption in Leggett's model is that purity of the state of each local subsystem should be retained. The marginal probabilities associated with local measurements performed on each of the subsystems should thus be compatible with such an assumption (they should be valid non-negative probability distributions). However, Leggett's model does not make any assumption on the joint correlations between different measurement outcomes on the two subsystems, thus explicitly allowing for a degree of non-locality ({\it i.e.}, Leggett's model can in general violate a Bell's inequality). The main point of Ref.~\cite{leggett} is that the compatibility requirements imposed on local marginals are strong enough to constrain even the non-local correlations. Quantum mechanics violates such constraints.}

{Various formulations of Leggett's original argument have been recently put forward and violation of non-local realism by polarization-encoded entangled states has been experimentally demonstrated in a series of seminal papers~\cite{vienna1,vienna2,singapore1,singapore2}. It is worth mentioning that, differently from the case of a Bell's inequality, where the bound imposed by local realistic theories does not depend on the measurement settings used in the actual implementation of the test, non-local realistic models enforce constraints that critically depend on the measurements being implemented. In the remainder of this paper we address two of such formulations and show that ECSs violate them up to the maximum allowed by a given configuration of measurement settings. In order to avoid unnecessary redundancies and technicalities, we refer to Refs.~\cite{vienna1,vienna2,singapore1,singapore2} for the full derivation of the inequalities that will be used here.}

\subsection{Resource state and tools}

\begin{figure}[t]
\centerline{\includegraphics[width=0.5\textwidth]{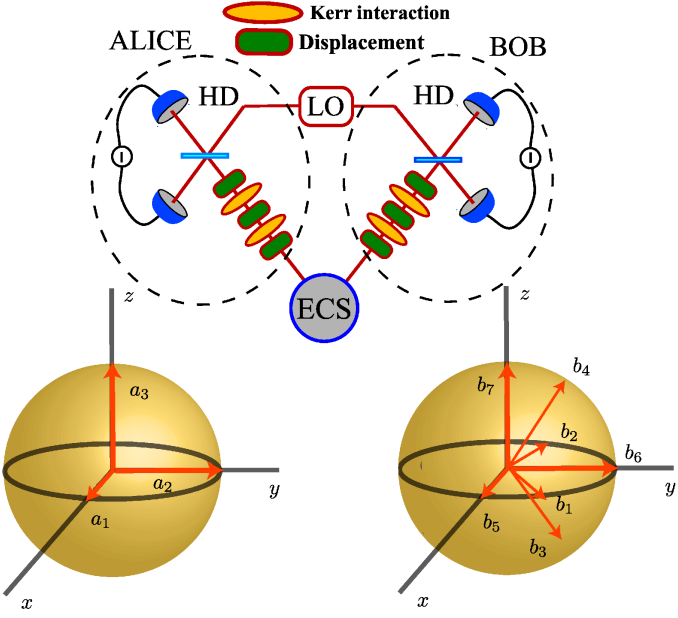}}
\caption{Scheme of the experiment for testing non-local realism with an entangled coherent state (ECS). The source generates an ECS of the form considered in the body of the paper. Alice and Bob perform local rotations through the sequence of unitary operations in Eq.~(\ref{sequence}). The locally-rotated  optical states are then mixed with a strong local oscillator (LO) and homodyne detectors (HDs) are used for final measurements. The leftmost and rightmost spheres show the directions of the vectors identifying the measurement settings at Alice's and Bob's sites respectively. }
\label{schema}
\end{figure}

{In this Subsection we formally introduce the class of CV states used in our analysis together with the formalism and tools necessary for  the measurements required by the Leggett tests at hand. Although bosonic modes of any nature could well be used in order to realize our proposal, it is natural to consider hereafter ECSs of optical field modes.} Among the states falling into the family of ECSs, we consider
\begin{equation}
\label{ecs}
\ket{\text{ECS}}_{AB}=\frac{\ket{\alpha,\alpha}_{AB}+\ket{-\alpha,-\alpha}_{AB}}{\sqrt{2(1+e^{-4|\alpha|^2})}},
\end{equation}
where $\ket{\alpha}=\hat{D}(\alpha)\ket{0}$ is a coherent state of amplitude amplitude $\alpha$, $\hat{D}(\alpha)=\exp[\alpha\hat{b}^\dag-\alpha^*\hat{b}]$ is the displacement operator, $\ket{0}$ is the vacuum state of a field mode with associated creation (annihilation) operator $\hat{b}^\dag$ ($\hat{b}$). In what follows, for easiness of calculation and without affecting the generality of our discussions, we consider only the case of $\alpha\in\mathbb{R}$. After generation of state~(\ref{ecs}), modes $A$ and $B$ are distributed to two agents, for definiteness called Alice and Bob, respectively. These have the task of performing local effective rotations and homodyne measurements over the respective subsystem. A sketch of such a thought experiment is shown in Fig.~\ref{schema}.

Differently from an optimized Bell-CHSH inequality, which requires 
measurement settings identified by vectors lying on the equatorial 
plane of a single-qubit Bloch sphere, Leggett's inequality needs the 
ability to perform out-of-plane measurements~\cite{vienna1}. This means that 
the following transformations should be realized  ($j{=}A,B$)
\begin{equation}
\label{rotazioni}
\begin{split}
&\ket{\alpha}_j\rightarrow\sin{\frac{\theta_j}{2}}\ket{\alpha}_j+e^{-i\varphi_j}\cos\frac{\theta_j}{2}\ket{-\alpha}_j,\\
&\ket{-\alpha}_j\rightarrow{e}^{i\varphi_j}\cos{\frac{\theta_j}{2}}\ket{\alpha}_j-\sin\frac{\theta_j}{2}\ket{-\alpha}_j.
\end{split}
\end{equation}
The $2{\times}{2}$ matrix describing Eq.~(\ref{rotazioni}) in the space spanned by $\{\ket{\alpha},\ket{-\alpha}\}$ can be decomposed into the sequence of elementary rotations $U_{z}(-\varphi_j/2)U_x(\pi/4)U_z(\vartheta_j/2)U_x(\pi/4)U_z(\varphi_j/2)$ with $U_{x,z}(\xi)=\text{Exp}[{i\xi\bm{\sigma}_{x,z}}]$ and where $\bm{\sigma}_{k}$ is the k-Pauli matrix ($k=x,y,z$). We now use the analysis performed in~\cite{jacobmyung}, where it is shown that the effect of $U_z(\xi)$ on a coherent state $\ket{\alpha}$ can be effectively approximated by a phase-space displacement operation $\hat{D}({i\xi}/{2\alpha})$, while $U_x(\pi/4)$ can be implemented by means of a proper Kerr-like single-mode nonlinearity $\hat{U}_{NL}=\text{Exp}[{-{i}\pi(\hat{a}^\dag\hat{a})^2}/2]$. Therefore, the physical implementation of Eqs.~(\ref{rotazioni}) would be given by the sequence 
\begin{equation}
\label{sequence}
{\hat{R}(\theta_j,\varphi_j)\!=\!\hat{D}_j(-i{\varphi_j}/{4\alpha})\hat{U}_{NL}\hat{D}_j
(i{\theta_j}/{4\alpha})\hat{U}_{NL}\hat{D}_j(i{\varphi_j}/{4\alpha})}.
\end{equation} 
From now on, the explicit form of Eqs.~(\ref{rotazioni}) will be specified by the directions of the unit vectors $\mathbf{a}\equiv(\theta_A,\varphi_A)$ and 
$\mathbf{b}\equiv(\theta_B,\varphi_B)$ identified by the corresponding 
set of angles expressed in spherical polar coordinates. After a lengthy but 
straightforward calculation, one gathers the explicit transformation experienced by $\ket{\pm\alpha}_j$
\begin{equation}
\label{esplicito}
\begin{split}
\ket{\alpha}_j&\rightarrow\frac{1}{2}\left\{e^{\frac{i\theta_j}{4}}\left[\vert{\alpha+\frac{i\theta_j}{4\alpha}}\rangle+ie^{\frac{i\varphi_j}{2}}\vert{-\alpha-\frac{i\varphi_j}{2\alpha}-\frac{i\theta_j}{4\alpha}}\rangle\right]\right.\\
&\left.+ie^{-\frac{i\theta_j}{4}}\left[e^{\frac{i\varphi_j}{2}}|-\alpha-\frac{i\varphi_j}{2\alpha}+\frac{i\theta_j}{4\alpha}\rangle+i\vert\alpha-\frac{i\theta_j}{4\alpha}\rangle\right]\right\},\\
\ket{-\alpha}_j&\rightarrow\frac{1}{2}\left\{ie^{\frac{i\theta_j}{4}}\left[i\vert{-\alpha-\frac{i\theta_j}{4\alpha}}\rangle+e^{-\frac{i\varphi_j}{2}}\vert{\alpha-\frac{i\varphi_j}{2\alpha}+\frac{i\theta_j}{4\alpha}}\rangle\right]\right.\\
&\left.+e^{-\frac{i\theta_j}{4}}\left[ie^{-\frac{i\varphi_j}{2}}|\alpha-\frac{i\varphi_j}{2\alpha}-\frac{i\theta_j}{4\alpha}\rangle+\vert-\alpha+\frac{i\theta_j}{4\alpha}\rangle\right]\right\}.
\end{split}
\end{equation}
These expressions are the starting point of our analysis. After the local transformations implemented by Alice and Bob, homodyne measurements are performed on system $A$ and $B$~\cite{barnettradmore}. These are arranged so that mode $A$ ($B$) is projected onto the in-phase quadrature eigenstate $\ket{x}$ ($\ket{y}$)~\cite{ioejacob,ioejacob2}. We can thus determine the joint probability-amplitude function
\begin{equation}
\label{prob}
C(\{\theta\},\{\varphi\},x,y)=_{AB}\!\langle{x,y}|\hat{R}(\theta_A,\varphi_A)\hat{R}(\theta_B,\varphi_B)\vert\text{ECS}\rangle_{AB},
\end{equation}
where $\{\theta\}\equiv\{\theta_A,\theta_B\}$ and $\{\varphi\}=\{\varphi_A,\varphi_B\}$ identify the two sets of relevant angles. 
For our test, we need a set of bounded dichotomic observables which we construct by assigning value $+1$ to the outcome of a homodyne measurement at Alice's (Bob's) site such that $x\ge{0}$ ($y\ge{0}$) and $-1$ otherwise. The joint probability of outcomes is thus written as 
\begin{equation}
\label{probjoint}
P_{kl}(\{\theta\},\{\varphi\})=\int^{k_s}_{k_i}\!d{x}\!\int^{l_s}_{l_i}dy|C(\{\theta\},\{\varphi\},x,y)|^2,
\end{equation}
where the subscripts $k,l=\pm$ correspond to Alice's and Bob's assignments of outcomes $\pm{1}$ and the integration limits are such that $+_s=\infty,+_i=-_s=0$ and $-_i=-\infty$. We can now introduce the correlation function
\begin{equation}
C^L(\{\theta\},\{\varphi\})=\sum_{k,l=\pm}P_{kk}(\{\theta\},\{\varphi\})-\sum_{k\neq{l}=\pm}P_{kl}(\{\theta\},\{\varphi\})
\end{equation}
which is needed in order to build up the proper Leggett-type function. The explicit calculation of $C^L(\{\theta\},\{\varphi\})$, performed by using the dependence of $\langle{x}|\alpha\rangle$ on Hermite polynomials and the Rodrigues formula~\cite{barnettradmore}, leads to 
\begin{widetext}
\begin{equation}
\label{esplicita}
\begin{split}
C^L(\{\theta\},\{\varphi\})&=\frac{e^{-({1}/{8\alpha^2}){\sum_{j=A,B}(8i\alpha^2+4\theta_j+\varphi_j)(4\theta_j+\varphi_j)}}}{32(1+e^{4\alpha^2})}\left\{8e^{{4\alpha^2+(1/8\alpha^2)\sum_{j=A,B}(8i\alpha^2+8\theta_j+\varphi_j)\varphi_j}{}}\prod_{j=A,B}[f_{-\theta_j,0}+e^{8i\theta_j}f_{\theta_j,0}]\right.\\
&-4e^{4\alpha^2}\prod_{j=A,B}[f_{-\theta_j,-\varphi_j}-e^{2\theta_j(4i+{\varphi_j}/{\alpha^2})}f_{\theta_j,-\varphi_j}]-4e^{4\alpha^2+2i(\varphi_A+\varphi_B)}\prod_{j=A,B}[e^{{2\theta_j\varphi_j}/{\alpha^2}}f_{-\theta_j,\varphi_j}-e^{i8\theta_j}f_{\theta_j,\varphi_j}]\\
&\left.+8e^{i\sum_{j=A,B}(4\theta_j+\varphi_j)}\prod_{j=A,B}[e^{{2\theta_j\varphi_j}/{\alpha^2}}g_{\theta_j,-\varphi_j}+g_{\theta_j,\varphi_j}]\right\}
\end{split}
\end{equation}
\end{widetext}
with $f_{\theta_j,\varphi_j}=\text{Erf}[{\sqrt{2}\alpha+i(4\theta_j+\varphi_j)}/{2\sqrt{2}\alpha}]$ and $g_{\theta_j,\varphi_j}=\text{Erfi}[({4\theta_j+\varphi_j})/{2\sqrt{2}\alpha}]$. This equation is the building block for the non-local realistic tests performed in Secs.~\ref{viennaformulation} and~\ref{singaporeformulation}.

\section{Leggett-type inequality violation}
\label{viennaformulation}

In this Section, we use an ECS resource for a Leggett-type test that does not require the impractical average of the correlation function over infinitely many measurement settings but, at the same time, does not rely on properties of rotational invariance of $C^L(\{\theta\},\{\varphi\})$, as instead required in~\cite{vienna1}. We make use of the simplest version of the class of inequalities discussed in~\cite{vienna2,singapore1}. 
Specifically, in Ref.~\cite{vienna2}, $C^L(\{\theta\},\{\varphi\})$ should be evaluated using seven pairs of bipartite measurement settings. We therefore introduce the unit vectors $\mathbf{a}\equiv(\theta_A,\varphi_A)$ 
and $\mathbf{b}\equiv(\theta_B,\varphi_B)$ specified by the set of corresponding  angles in spherical polar coordinates.
\begin{figure}[b]
\centerline{\includegraphics[width=0.45\textwidth]{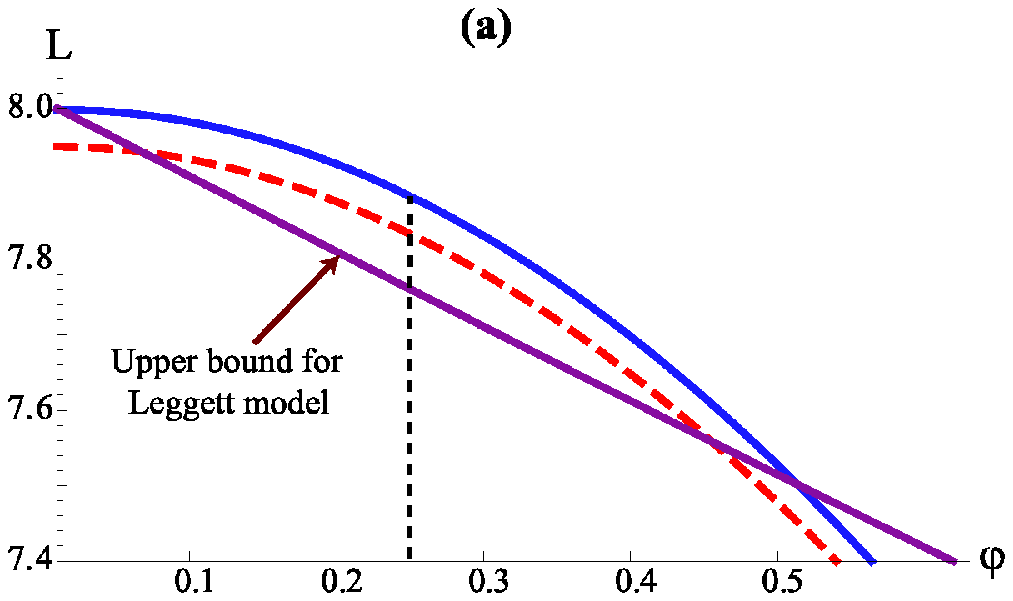}}
\vspace{0.5cm}
\centerline{\includegraphics[width=0.45\textwidth]{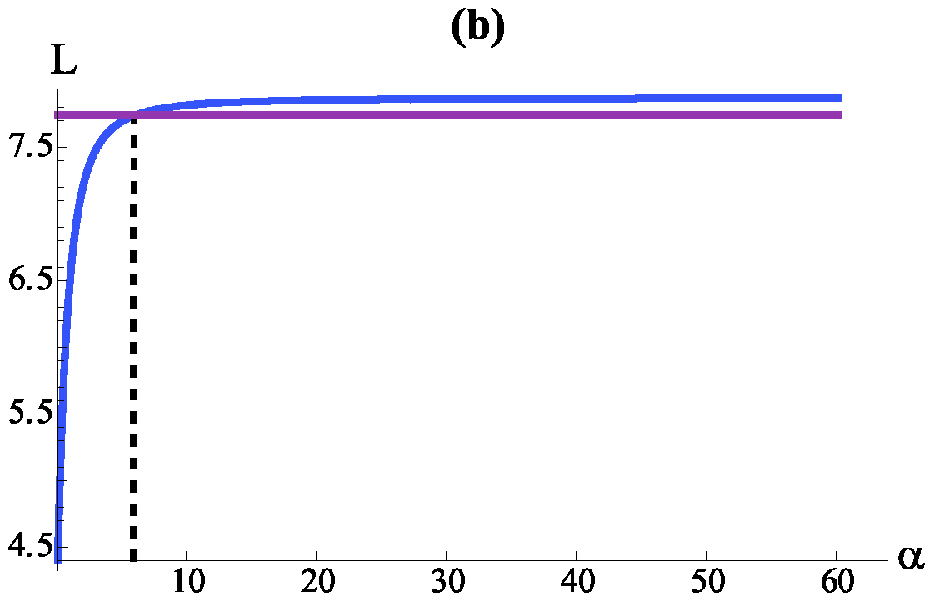}}
\caption{{\bf (a)} Violation of non-local realism by the Leggett function $L_{}$ in Eq.~(\ref{leggettine}) for ${\alpha=10}$ (dashed curve) and $\alpha=60$ (solid curve). The upper bound for Leggett model is also plotted against the angle $\varphi$. The vertical dashed line indicates the value $\varphi\simeq{0.2507}$ radians at which the Leggett-type inequality in~\cite{singapore2} is maximally violated, regardless of $\alpha$. {\bf (b)} Violation of Leggett's model by $L$ against the amplitude of the coherent state component in the ECS considered in the body of the paper. In this plot we have assumed $\varphi=0.25$. The upped bound for Leggett model (straight line) is surpassed for $\alpha\ge7.5$.}
\label{disuVIENNA}
\end{figure}
A Leggett-type inequality can now be tested by considering the unit vectors 
$\mathbf{a}_{1,2,3}$ and $\mathbf{b}_{1-7}$, each 
specifying a rotation that Alice (Bob) should perform on her (his) mode. 
Explicitly 
\begin{equation}
\begin{split}
\mathbf{a}_{1}&=\mathbf{b}_{5}\equiv({\pi}/{2},0),~\mathbf{a}_{2}=\mathbf{b}_{6}\equiv({\pi}/{2},{\pi}/{2}),\\
\mathbf{a}_{3}&=\mathbf{b}_{7}\equiv({0}{},{0}{}),~\mathbf{b}_{1}\equiv({\pi}/{2},\varphi),~\mathbf{b}_{4}\equiv({\varphi},{\pi}/{2})
\end{split}
\end{equation}
 with $\mathbf{b}_{2}$ and $\mathbf{b}_{3}$ which are found from $\mathbf{b}_{1}$ and $\mathbf{b}_{4}$, respectively, by taking 
$\varphi\rightarrow\pi/2+\varphi$. These vectors are clearly represented in the Bloch spheres of Fig.~\ref{schema}.
With these definitions, we consider the Leggett function~\cite{vienna2,singapore1}
\begin{equation} 
\label{leggettine} 
\begin{split} 
&L\!=\!|C^L_{}(\mathbf{a}_1,\mathbf{b}_1)\!+\!C^L_{}
(\mathbf{a}_2,\mathbf{b}_2)\!+\!C^L_{}
(\mathbf{a}_1,\mathbf{b}_5)\!+\!C^L_{}(\mathbf{a}_2,\mathbf{b}_6)|\\ 
&\!+\!|C^L_{}(\mathbf{a}_2,\mathbf{b}_3)\!+\!C^L_{}
(\mathbf{a}_3,\mathbf{b}_4)\!+\!C^L_{}(\mathbf{a}_2,\mathbf{b}_6)\!+\!C^L_{}(\mathbf{a}_3,\mathbf{b}_7)|. 
\end{split} 
\end{equation}
In contrast with a Bell-CHSH test, Leggett's non-local realistic theory imposes a bound on $L$ which actually depends on the relative direction of the measurement-setting vectors. Specifically, the inequality that non-local realistic models should satisfy reads $L\le8-2|\sin(\varphi/2)|$~\cite{vienna2,singapore1}. In what follows we show that an ECS of sufficiently large amplitude $\alpha$ always violates this constraint. 

Both the bound and the Leggett function have been plotted in Fig.~\ref{disuVIENNA} {\bf (a)} against the angle $\varphi$. In analogy with what happens in a Bell-CHSH test on ECS performed with homodyne measurements~\cite{stobinska}, we expect the dependence of the Leggett function on the amplitude $\alpha$ of the ECS resource. In fact, both the degree of violation and  the values of $\varphi$ such that $L$ is larger than the corresponding bound depend on $\alpha$, as shown in Fig.~(\ref{disuVIENNA}) {\bf (a)}, where the cases of $\alpha{=}10$ and $60$ are presented. On the other hand, the value of $\varphi$ maximizing the discrepancy with the non-local realistic theory is insensitive to the amplitude of the coherent states. Numerically, we have found that the function ${\cal L}=L-8+|\sin(\varphi/2)|$ that measures the degree of violation of such non-local realistic model is maximized for $\varphi\simeq{0.25}$ radians, which is the value we retain in our calculations. For this choice of $\varphi$, Fig.~\ref{disuVIENNA} {\bf (b)} reveals that the maximum degree of violation is achieved quite quickly as $\alpha$ grows. For $\alpha\gtrsim{10}$, a quasi-{\it plateau} is achieved close to $L\sim{7.87}$, which is in excellent agreement with the expected value of $L$, at $\varphi=0.25$, for a pure singlet state~\cite{vienna2,singapore2}. This demonstrates that non-local realistic models should be abandoned for an ECS of large enough amplitude. Even modest values of $\alpha$ allow for the maximum violation of such Leggett-type inequality, therefore mimicking the results expected and observed for the singlet state of two qubits. 

\begin{figure}[t]
\centerline{\includegraphics[width=0.4\textwidth]{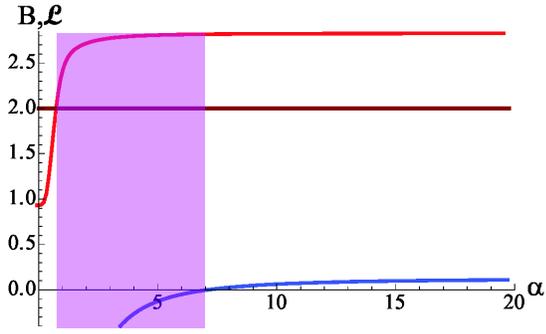}}
\caption{Violation of local and non-local realism against the amplitude $\alpha$ of the ECS components. We show the Bell function $B$ (together with the local realistic bound $2$) and the function ${\cal L}$, which violates the Leggett-type inequality when positive. The shaded region corresponds to values of $\alpha$ where local realistic theories should be abandoned while Leggett's inequality is still satisfied.}
\label{disuVIENNAcomp}
\end{figure}

As already stated, Leggett's model assumes that the state of the local elements of a bipartite state is pure. 
The interplay between local and non-local realism in the space of parameters of a given bipartite state is an issue not yet fully explored~\cite{ioejacob}. Here, we perform a step in this direction by comparing the behavior of $L$ and the Bell-CHSH function obtained by using an ECS, local rotations and homodyne measurements~\cite{stobinska,commento1}. Fig.~\ref{disuVIENNAcomp} shows Leggett's and Bell-CHSH functions, numerically optimized over the corresponding measurement settings, against $\alpha$. The Bell-CHSH inequality is violated already for $\alpha\gtrsim{1}$ while, as seen in Fig.~\ref{disuVIENNA} {\bf (b)}, $\alpha$ should be increased up to $7$ in order to violate the Leggett-type inequality we are studying. The existence of an ample region where ${\cal L}\leq 0$ while local realism should be abandoned is interesting. Although, clearly, no firm statement can be drawn, it is tempting to ``retain" non-local realistic theories to explain all the measurement results under our assumptions, in such region, a point which has been discussed in detail in Ref.~\cite{ioejacob}.

\section{Optimal Leggett-type inequality}
\label{singaporeformulation}

Very recently, Branciard {\it et al.} have proposed and experimentally 
tested a new family of Leggett-type inequality which supersedes those presented  in~\cite{vienna2,singapore1} in terms of number of required measurement settings at Bob's site. The only assumption in Branciard {\it et al.}'s derivation is the existence of valid conditional probability distribution for the outcomes of the measurements performed by Alice and Bob~\cite{singapore2}. The simplest inequality that can be derived in this context needs the Leggett function
\begin{equation}
\label{ineqSingapore}
L_{S}=({1}/{3})\sum^3_{i=1}\vert{C}^L({\bf a}_i,{\bf b}^+_{i})+{C}^L({\bf a}_i,{\bf b}^-_{i})\vert
\end{equation}
and reads $L_S\le{2-(2/3)|\sin(\varphi/2)|}$. The number of measurement settings required at Bob's site for this test is only 6. While ${\bf a}_{i}$'s ($i=1,2,3$) coincide with those used in order to build Eq.~(\ref{leggettine}), we have 
\begin{equation}
{\bf b}^\pm_{1}{\equiv}({\pi}/{2},\pm{\varphi}/{2}),~
{\bf b}^\pm_{2}{\equiv}({\pi}/{2}\mp{\varphi}/{2},{\pi}/{2}),~
{\bf b}^\pm_{3}{\equiv}(\pm{\varphi}/{2},0).
\end{equation}
\begin{figure}[t]
\centerline{\includegraphics[width=0.45\textwidth]{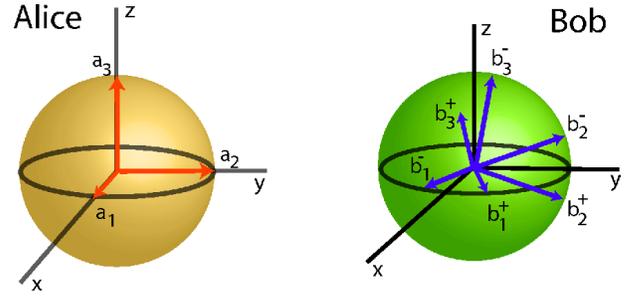}}
\caption{Measurement settings at Alice and Bob's site for the test of non-local realism as proposed by Branciad {\it et al.} in Ref.~\cite{singapore2}. The pairs of vectors $({\bf b}_{i},{\bf b}'_{i})$ with $i=1,2,3$ lie on orthogonal planes, the vectors forming angles equal to $\varphi$. The pair with $i=1,2,3$ lies on the plane with ${z,x,y}=0$, respectively.}
\label{sfere1}
\end{figure}
\begin{figure}[b]
\centerline{\includegraphics[width=0.4\textwidth]{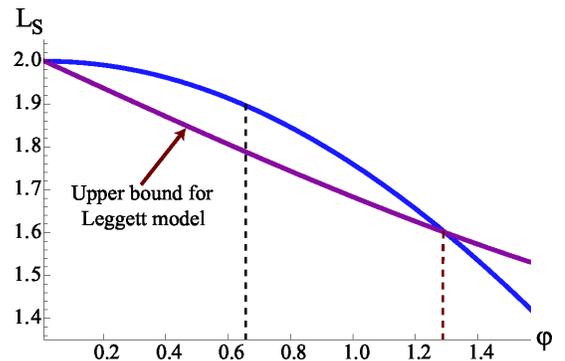}}
\caption{Violation of a Leggett-type inequality by the function $L_{S}$ in Eq.~(\ref{ineqSingapore}) for $\alpha=60$. The upper bound for Leggett model is also plotted against the angle $\varphi$. The leftmost vertical dashed line indicates the value $\varphi\simeq{0.65}$ radians at which the Leggett-type inequality in~\cite{singapore2} is maximally violated. The rightmost one, corresponding to $\varphi\simeq{1.28}$ radians, sets the upper bound for $\varphi\in[0,\pi/2]$ for the Legget-type inequality.}
\label{disuSINGAPORE}
\end{figure}
As it is also shown in Fig.~\ref{sfere1}, the pairs of measurement-setting vectors $({\bf b}^+_i,{\bf b}^-_i)$ lie on orthogonal planes and form an angle $\varphi$. By following the discussion in Sec.~\ref{viennaformulation}, it is straightforward to check the violation of non-local realism  by an ECS. The results are shown for $\alpha=60$ in Fig.~\ref{disuSINGAPORE}, where at $\varphi_{max}\simeq{0.65}$ the maximal violation of Leggett's model is achieved. At this value, while the local realistic bound equals $\simeq1.787$, we have $L_{S}\simeq{1.898}$. Both this value and $\varphi_{max}$ are in excellent agreement with the expectations for the discrete-variable case~\cite{singapore2}. As before, the degree of violation depends on the amplitude of the coherent state components used in the ECS resource. A picture analogous to the one presented in Figs.~\ref{disuVIENNA} can be easily drawn. We omit it here for the sake of conciseness.

\section{Effects of detection inefficiency}
\label{inefficiencies}

In order to include the effects of non-ideal efficiency of the homodyne detectors, we need to modify our approach.
An imperfect homodyne detector with efficiency $\eta$
can be modelled by a beam splitter with transmittivity $\eta$ superimposing modes $j=A,B$
with an ancillary mode $a_j$ prepared in vacuum state and cascaded with a perfect homodyne detector.
In this way, part of the field that should arrive at the perfect homodyne detector is
tapped by the beam splitter. The beam splitter operation between
modes $j$ and $a_j$ is defined as
${\hat B}_{ja_j}=\exp[{{\zeta}{}({\hat b}_j^\dagger {\hat b}_{a_j}
-{\hat b}_j {\hat b}_{a_j}^\dagger)}/2]$,
where $\cos\zeta=\sqrt{\eta}$. Via the dichotomization process described in Sec.~\ref{tools}, the correlations entering a Leggett function can also be expressed as
\begin{equation}
C^L_{d}(\{\theta\},\{\varphi\})\!=\!
\int{d}x_{Ar}{d}y_{Br}\text{s}(x_{r}y_{r}){\cal P}(\{\theta\},\{\varphi\},x_{r},y_{r})
\end{equation}
where $x=x_{r}+ix_{i}$ and $y=y_{r}+iy_{i}$ are the complex in-phase quadrature variables
 and ${\cal P}(\{\theta\},\{\varphi\},x_{r},y_{r})$ is a marginal probability distribution calculated
  from the total Wigner function of modes $A$ and $B$ after the trace over the ancillae.

\begin{figure*}[t]
\centerline{{\bf (a)}\hskip9cm{\bf (b)}}
\centerline{\includegraphics[width=0.4\textwidth]{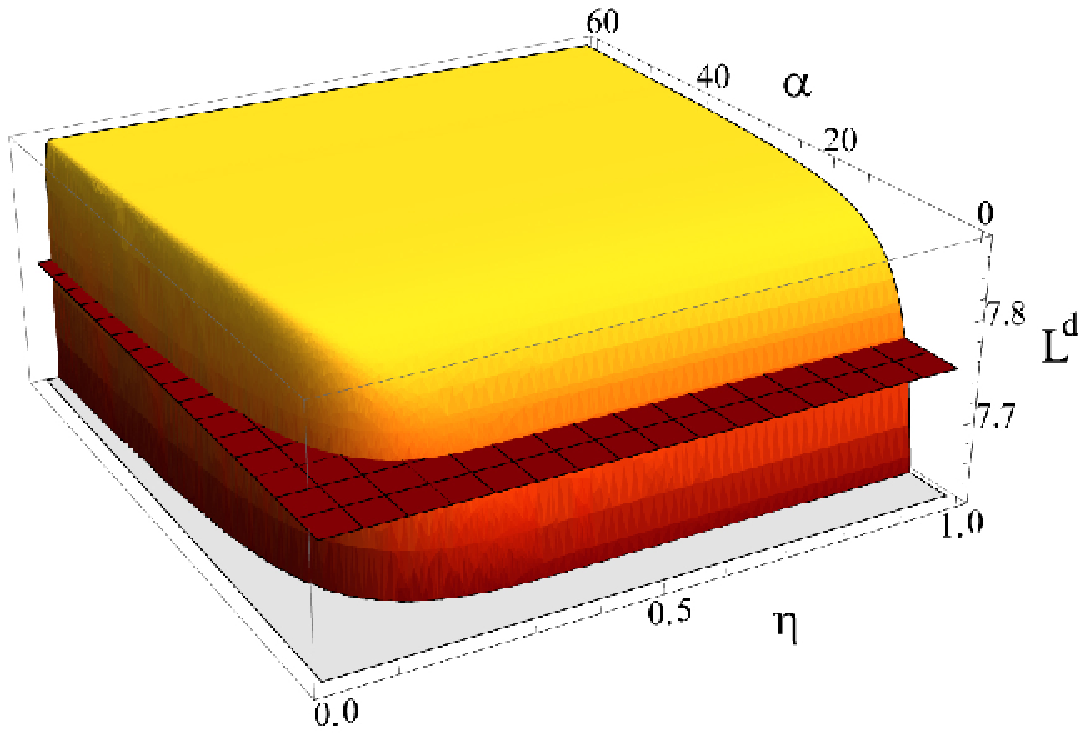}\includegraphics[width=0.45\textwidth]{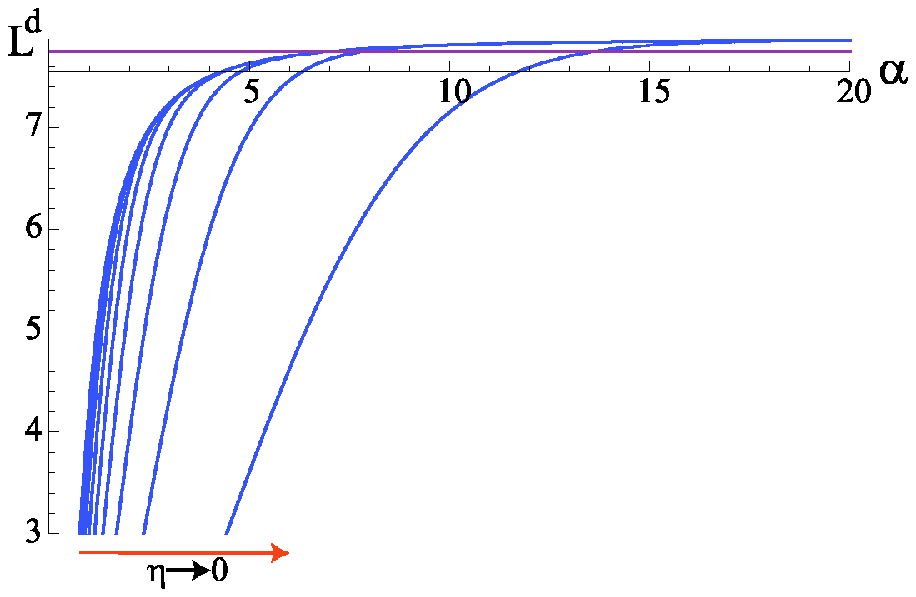}}
\caption{{\bf (a)} Leggett function with inefficient detection $L^d$ plotted against the coherent state amplitude $\alpha$ and the detection inefficiency $\eta$. The horizontal plane is the non-local realistic bound for the inequality in Refs.~\cite{vienna2,singapore1} at $\varphi\simeq{0.25}$. {\bf (b)} $L^d$ versus $\alpha$ for decreasing values of $\eta$, which goes from $1$ to $0.1$ (in steps of $0.1$) in going from the leftmost to the rightmost curve. The straight line indicates the non-local realistic bound.}
\label{disuVIENNAEfficiency}
\end{figure*}
   The calculation of the latter is sketched as follow. First, we determine the Weyl 
   characteristic function of state $\hat{R}(\theta_A,\varphi_A)\hat{R}(\theta_B,\varphi_B)\ket{\text{ECS}}_{AB}$, which reads 
\begin{equation}
\label{weyl}
\chi=_{AB}\!\langle{\text{ECS}}|\hat{\cal D}_A(\mu_A,\theta_A,\varphi_A)\hat{\cal D}_B(\mu_B,\theta_B,\varphi_B)\ket{\text{ECS}}_{AB}.
\end{equation}
This equation shows that $\chi$ is 
the sum of matrix elements (over coherent states) 
of {\it rotated displacement operators}  $\hat{\cal D}(\mu_j,\theta_j,\varphi_j)=\hat{R}^{\dag}(\theta_j,\varphi_j)\hat{D}_{j}(\mu_j)\hat{R}^{}_{j}(\theta_j,\varphi_j)$, 
 each being very easily evaluated using the operator-expansion formula~\cite{barnettradmore}, Eqs.~(\ref{rotazioni}) 
 and the relation
\begin{equation}
\langle\sigma|\hat{D}_j(\mu_j)|\tau\rangle=e^{-\frac{1}{2}
(|\sigma|^2+|\mu_j|^2+|\tau|^2)+\sigma^*\mu_j+\tau(\sigma^*-\mu^*_j)},
\end{equation}
where $\ket{\sigma}$ and $\ket{\tau}$ are arbitrary coherent states. 
The Wigner function is then calculated through the Fourier transform
 of $\chi$ as
\begin{equation}
W_{\{\theta\},\{\varphi\}}(x_A,y_B)\!=\!\frac{1}{\pi^4}\int{d}^2
{\mu_A}d^2\mu_B\chi{e}^{x^*_A\mu_A+y^*_B\mu_B-h.c.}.
\end{equation}
While the calculation is straightforward, the explicit form of this function
is rather uninformative and we omit it. The effects of detection inefficiencies are now included by convoluting $W_{\{\theta\},\{\varphi\}}(x_A,y_B)$
with the Wigner function of two ancillary modes prepared in their vacuum state and considering the action of the beam splitters used to model the inefficient detectors on the quadrature variables.
We call
$W^d_{\{\theta\},\{\varphi\}}(x_A,y_B,\eta)$ the Wigner function of the reduced state of 
$A$ and $B$ after the degrees of freedom of the ancillae are integrated out. From this,
the marginal probability distribution is extracted as
\begin{equation}
\label{marginal}
{\cal P}(\{\theta\},\{\varphi\},x_{r},y_{r})=\int{d}{x}_{i}dy_{i}{W}^d_{\{\theta\},\{\varphi\}}(x_A,y_B,\eta),
\end{equation}
which is all we need in order to get $C^L_d(\{\theta\},\{\varphi\})$. With this, the Leggett function $L^d$ is found as in Eq.~(\ref{leggettine}) by replacing $C^L(\{\theta\},\{\varphi\})$ with $C^L_d(\{\theta\},\{\varphi\})$ and the inequality discussed in Sec.~\ref{viennaformulation} can be studied. The results are shown in Figs.~\ref{disuVIENNAEfficiency}. The value of $\varphi$ which maximizes the inequality violation is independent of $\eta$, which is kept as $0.25$ throughout this Section. The effects of decreasing detectors efficiencies amounts in increasing the threshold values of $\alpha$ at which ${\cal L}^d=L^d-8+2|\sin(\varphi/2)|$ becomes positive. This is clearly seen in Fig.~\ref{disuVIENNAEfficiency} {\bf (b)} where it is shown that, even with extremely inefficient detectors, a sufficiently large value of $\alpha$ allows for maximal violation of non-local realism, a feature that is unique of the proposed test for non-local realism based on the use of ECS resources.

\section{Conclusions}
\label{conclusions}

We have investigated the violation of non-local realism using ECSs, local rotations implemented by nonlinear media and inefficient homodyne measurements. Our study reveals that, by reducing the overlap between the components of the ECS used to test non-local realism, therefore faithfully mimiking a two-qubit state, violations of an optimal Leggett inequality up to the maximum allowed value is achieved. 

{Our work contributes to the characterization of the properties of ECSs as resources having important and intriguing applications in quantum technology. The fact that ECSs allow for the violation of non-local realism is an accomplishment that should be valued alongside the violation of BellÕs and Mermin-Klysko inequalities by this very same class of states~\cite{stobinska}. On one hand, our study enlarges the range of useful and interesting applications of the class of entangled states embodied by ECSs. On the other hand, we believe our work is endowed with further relevance, as it provides the recipe for the implementation of all the necessary steps in the Leggett's test at hand and relies on experimentally non-demanding homodyne measurements.} A demonstration of our predictions may be realized by generating the required ECS using a beam splitter, one input mode in the vacuum state and the other prepared in a superpositions of two coherent states as recently implemented in Ref.~\cite{jacobnature}. The resource state is therefore not far-fetched. On the other hand, it is clear from our analysis that an important role is played by the nonlinear dynamics at the basis of the effective local rotations used for the Leggett test. The crucial point, here, would be the achievement of a large enough nonlinear rate. Very important progresses have been made in this direction~\cite{nonlinear} and one can be confident that the tchnological gap will soon be filled. Our work contributes to current studies on the interplay between locality and realism by proposing a novel scenario where such a trade off, which is crucial in the context of modern quantum mechanics, can be quantitatively analized. 

\acknowledgments

The authors thank T. C. Ralph for stimulating discussions and comments. MP thanks M. S. Kim and T. Paterek for useful discussions. This work is supported by the UK EPSRC (EP/G004579/1), the Center for Theoretical Physics at Seoul National University, the World Class University (WCU) program and the KOSEF grant funded by the Korea government (MEST) (R11-2008-095-01000-0).


\begin {thebibliography}{99}

\bibitem{horo} M. A. Nielsen and I. L. Chuang, {\it Quantum Computation and Quantum Information} (Cambridge University Press, Cambridge, 2000); A. Peres, {\it Quantum Theory, Concepts and Methods} (Kluwer, 1993); L. Amico, R. Fazio, A. Osterloh, and V. Vedral, \rmp {\bf 80}, 517 (2008); R. Horodecki, P. Horodecki, M. Horodecki, and K. Horodecki, Rev. Mod. Phys. {\bf 81}, 865 (2009).

\bibitem{bell} J. S. Bell, Physics {\bf 1}, 195 (1964); J. S. Bell, {\it Speakable and Unspeakable in Quantum Mechanics}, (Cambridge University Press, Cambridge, 1987).

\bibitem{exp} A. Aspect, J. Dalibard, and G. Roger, \prl {\bf 49}, 1804 (1982); G. Weihs, T. Jennewein, C. Simon, H. Weinfurter, and A. Zeilinger, \prl {\bf 81}, 5039 (1998).

\bibitem{leggett} A. Leggett, Found. Phys. {\bf 33}, 1469 (2003).

\bibitem{vienna1} S. Gr\"oblacher, T. Paterek, R. Kaltenbaek, C. Brukner, M. Zukowski, M. Aspelmeyer, and A. Zeilinger, Nature (London), {\bf 446}, 871 (2007).

\bibitem{vienna2} T. Paterek, A. Fedrizzi, S. Gr\"oblacher, T. Jennewein, M. Zukowski, M. Aspelmeyer, and A. Zeilinger, \prl {\bf 99}, 210406 (2007).

\bibitem{singapore1} C. Branciard, A. Ling, N. Gisin, C. Kurtsiefer, A. Lamas-Linares, and V. Scarani, \prl {\bf 99}, 210407 (2007).

\bibitem{singapore2} C. Branciard, N. Brunner, N. Gisin, C. Kurtsiefer, A. Lamas-Linares, A. Ling, and V. Scarani, Nature Phys. {\bf 4}, 681 (2008).

\bibitem{migdall} M. D. Eisaman, E. A. Goldschmidt, J. Chen, J. Fan, and A. Migdall, \pra {\bf 77}, 032339 (2008).

\bibitem{bw} K. Banaszek and K. W\'odkiewicz, \pra {\bf 58}, 4345 (1998); \prl {\bf 82}, 2009 (1999).

\bibitem{chen} Z. Chen, J. Pan, G. Hou, Y. Zhang, \prl {\bf 88}, 040406 (2002).

\bibitem{jacobmyungetal} H. Jeong, W. Son, M. S. Kim, D. Ahn, and C. Brukner, \pra {\bf 67}, 012106 (2003).

\bibitem{jacobralph} H. Jeong and T. C. Ralph, Phys. Rev. Lett. {\bf 97}, 100401 (2006).

\bibitem{ioejacob} H. Jeong, M. Paternostro, and T. C. Ralph, Phys. Rev. Lett. {\bf 102}, 060403 (2009).

\bibitem{ioejacob2} M. Paternostro, H. Jeong, and T. C. Ralph, Phys. Rev. A {\bf 79}, 012101 (2009).

\bibitem{ECS} B. C. Sanders, \pra {\bf 45}, 6811 (1992); B. C. Sanders, K. S. Lee, and M. S. Kim, \pra {\bf 52}, 735 (1995). 

\bibitem{examples} P. T. Cochrane, G. J. Milburn, and W. J. Munro, Phys. Rev. A {\bf 59}, 2631 (1999); S. J. van Enk and O. Hirota, \pra {\bf 64}, 022313 (2001); X. Wang, \pra {\bf 64}, 022302 (2201); H. Jeong, M. S. Kim, and J. Lee, \pra {\bf 64}, 052308 (2001); W. J. Munro, K. Nemoto, G. J. Milburn, and S. L. Braunstein, Phys. Rev. A {\bf 66}, 023819 (2002); T. C. Ralph, A. Gilchrist, G. J. Milburn, W. J. Munro, and S. Glancy, Phys. Rev. A {\bf 68}, 042319 (2003); H. Jeong and M. S. Kim, Quant. Inf. Comp. {\bf 2}, 208 (2002); Nguyen Ba An, Phys. Rev. A {\bf 68}, 022321 (2003); Phys. Rev. A {\bf 69}, 022315 (2004); M. Paternostro, M. S. Kim, and P. L. Knight, \pra {\bf 71}, 022311 (2005); A. P. Lund, T. C. Ralph, and H. L. Haselgrove, Phys. Rev. Lett. {\bf 100}, 030503 (2008).

\bibitem{jacobmyung} H. Jeong and M. S. Kim, \pra {\bf 65}, 042305 (2002).

\bibitem{derek} D. Wilson, H. Jeong, and M. S. Kim, J. Mod. Opt. {\bf 49}, 851 (2002).

\bibitem{stobinska} M. Stobi\'nska, H. Jeong, and T. C. Ralph, \pra {\bf 75}, 052105 (2007); H. Jeong and Nguyen Ba An, Phys. Rev. A {\bf 74}, 022104 (2006).

\bibitem{commento1} Incidentally, Fig.~\ref{disuVIENNAcomp} reproduces Fig.~3 {\bf (a)} in Ref.~\cite{ioejacob}, where the case of a mixed bipartite entangled state was addressed.
 
\bibitem{barnettradmore} S. M. Barnett and P. M. Radmore, {\it Methods in Theoretical Quantum Optics} (Clarendon Press, Oxford, 1997).

\bibitem{jacobnature} A. Ourjoumtsev, H. Jeong, R. Tualle-Brouri, and P. Grangier, Nature {\bf 448}, 784 (2007). 

\bibitem{nonlinear} L. V. Hau, S. E. Harris, Z. Dutton, and C. H. Behroozi, Nature (London) {\bf 397}, 594 (1999); P. Bermel, A. Rodriguez, J. D. Joannopoulos, and M. Soljacic, Phys. Rev. Lett. {\bf 99}, 053601 (2007); F. G. S. L. Brandao, M. J. Hartmann, and M. B. Plenio, New J. Phys. {\bf 10}, 043010 (2008).



\end {thebibliography}
\end{document}